# Using FU Orionis Outbursts to Constrain Self–Regulated Protostellar Disk Models[1]

K. R. Bell and D. N. C. Lin[2]

University of California Observatories / Lick Observatory

University of California, Santa Cruz, CA 95064

## ABSTRACT

One dimensional, convective, vertical structure models and one dimensional, time dependent, radial diffusion models are combined to create a self–consistent picture in which FU Orionis outbursts occur in young stellar objects (YSOs) as the result of a large scale, self–regulated, thermal ionization instability in the surrounding protostellar accretion disk. Although active accretion disks have long been postulated to be ubiquitous among low mass young stellar objects, few constraints have until now been imposed on physical conditions in these disks. By fitting the results of time dependent disk models to observed time scales of FU Orionis events, we estimate the magnitude of the effective viscous stress in the inner disk ($r \lesssim 1$ AU) to be, in accordance with an *ad hoc* "alpha" prescription, the product of the local sound speed, pressure scale height, and an efficiency factor $\alpha$ of $10^{-4}$ where hydrogen is neutral and $10^{-3}$ where hydrogen is ionized.

We hypothesize that all YSOs receive infall onto their outer disks which is steady (or slowly declining with time) and that FU Orionis outbursts are self-regulated, disk outbursts which occur *only* in systems which transport matter inward at a rate sufficiently high to cause hydrogen to be ionized in the inner disk. We estimate a critical mass flux of $\dot{M}_{crit} = 5 \times 10^{-7}$ M$_\odot$/ yr *independent of the magnitude of $\alpha$* for systems with one solar mass, three solar radius central objects. Infall accretion rates in the range of $\dot{M}_{in} = (1 - 10) \times 10^{-6}$ M$_\odot$/ yr produce observed FU Orionis time scales consistent with estimates of spherical molecular cloud core collapse rates. Modeled ionization fronts are typically initiated near the inner edge of the disk and propagate out to a distance of

---

several tens of stellar radii. Beyond this region, the disk transports mass steadily inward at the supplied constant infall rate. Mass flowing through the innermost disk annulus is equal to $\dot{M}_{in}$ only in a time averaged sense and is regulated by the ionization of hydrogen in the inner disk such that long intervals ($\approx 1000$ yrs) of low mass flux: $(1 - 30) \times 10^{-8}$ $M_{\odot}/$ yr, are punctuated by short intervals ($\approx 100$ yrs) of high mass flux: $(1 - 30) \times 10^{-5}$ $M_{\odot}/$ yr. Time scales and mass fluxes derived for quiescent and outburst stages are consistent with estimates from observations of T Tauri and FU Orionis systems respectively.

*Subject headings:* accretion disks: protostellar, stars: pre- Main Sequence

## 1. INTRODUCTION

Disks are thought to exist around 30–50% of all low mass ($0.5$–$3$ $M_{\odot}$) young stellar objects (Strom, Edwards, & Strutskie 1993). That the disks are *actively accreting* rather than merely passively reradiating light from the central object is evidenced by the correlation of the existence of infrared excess with the existence of ultraviolet excess (Bertout & Bouvier 1989). T Tauri accretion rates inferred from UV excesses are generally in the range of $10^{-8}$ to a few $10^{-7}$ $M_{\odot}/$ yr (Bertout, Basri, & Bouvier 1988). Until recently, protostellar mass buildup was thought to be a quasi steady phenomenon in which infall, occurring first spherically from a molecular cloud core and later equatorially from an accretion disk, declined smoothly. Over the past decade, however, the quasi–static nature of the accretion disk phase has been challenged by successful modeling of observed large scale luminosity fluctuations as changes in accretion rates through surrounding protostellar disks (cf. Hartmann, Kenyon, & Hartigan 1993). Meteoritic evidence (Wood 1985) suggests frequent eruptive events may also have occurred in the solar nebula. We use a particularly dramatic form of outburst: the FU Orionis event, to put constraints on conditions in protostellar accretion disks.

Because YSO protostellar disks are optically thick out to many AU (Strom et al. 1993), conditions at the midplane of accretion disks in the planet forming region are not directly observable and can only be derived through the use of theoretical modeling. The primary and crucial unknown in disk calculations has historically been the mechanism by which angular momentum is transported outward and thus by which mass flows inward (Lynden-Bell & Pringle 1974). Among the many proposed angular momentum transport mechanisms, viscous stress is considered to be one of the most likely processes (Pringle 1981). However, the magnitude of simple molecular viscosity is too small to induce protostellar disks to evolve on reasonable time scales (von Weizsäcker 1943). It is customary to assume that a turbulent viscosity allows the mixing of adjacent annuli which results in a net outward transport of angular momentum and inward transport of matter and a consequent local generation of energy (Lüst 1952). Possible causes for turbulence in protostellar disks include convective instability (Cameron 1969; Lin & Papaloizou 1980) and magnetohydrodynamic instability (Balbus & Hawley 1991). The most commonly used model for turbulent viscosity in accretion disks is the *alpha* ($\alpha$) prescription (Shakura & Sunyaev 1973). This approximation is particularly useful for parameter studies of disk structure and evolution without exact specification of the nature of the effective viscosity. The main drawbacks to this model are its *ad hoc* functional form and the considerable uncertainty of the efficiency factor $\alpha$.

One of the main successes of the $\alpha$ prescription is its use in explaining the episodic 2–5 magnitude outbursts, on the time scale of weeks to months, in cataclysmic variables (hereafter CVs) (Warner 1976; Smak 1984). These systems are close binary stars in which a main sequence dwarf transfers its mass through the process of Roche lobe overflow to a white dwarf companion via a geometrically thin accretion disk (Bath et al. 1974; Smak 1983). For typical mass transfer rates in CVs (Patterson 1983), some regions of the disk are expected to be partially ionized. A generally accepted scenario for the outburst mechanism is a disk driven thermal instability induced by $H^-$ opacity in partially ionized disk regions (Meyer & Meyer-Hofmeister 1981; Smak 1984; Faulkner, Lin, & Papaloizou 1983; Mineshige & Osaki 1983, 1985; Cannizzo & Wheeler 1993). Provided the effective viscosity increases with temperature, as in the case of the $\alpha$ prescription, mass transfer in the disk may be modulated by the thermal instability in such a way as to reproduce observed CV outbursts (Papaloizou, Faulkner, & Lin 1983). Numerical computations on the dynamical evolution of thermally unstable disks indicate that the detailed features in the light curves are essentially determined by the magnitude of effective viscosity (Lin, Papaloizou, & Faulkner 1985).

Based on the similarities between the light curves of FU Orionis events (Herbig 1977) and those of CVs, a scenario was proposed identifying thermal instability in protostellar disks as the cause of FU Orionis events (Hartmann & Kenyon 1985; Lin & Papaloizou 1985). It is noted that the interpretation of the FU Orionis outburst as an accretion event is not universally accepted. An alternative explanation in which features are explained as arising in a rotating stellar envelope has been supported by Herbig (1989) and Petrov & Herbig (1992). We also mention the work of Stahler (1989) in which outbursts occur due to a relaxation of the stellar envelope, but agree with his assessment that the once per star hypothesis is not justified by the statistics (§2.4. below and Herbig 1977, 1989). While the accretion event hypothesis may not be an entirely unique interpretation of the data, it is one which fits the bulk of the evidence.

Throughout this paper we assume the thermally regulated accretion event hypothesis as a motivation for our outburst models and make use of the observed light curves of FU Orionis events to constrain the magnitude of protostellar disk viscosity. While small scale variability may be due to local clumpiness or feedback effects in the optically thin T Tauri boundary layer, large scale outbursts which are regulated by the disk's viscosity propagate at large scale thermal or viscous time scales. Studying such global outbursts can reveal important clues to the nature and magnitude of viscosity which in turn can result in detailed information about midplane conditions. In §2. we discuss the time scales and outburst properties of FU Orionis objects. In §3. we briefly recapitulate the development of accretion disk theory which we use in a series of numerical models (§4.) designed to mimic FU Orionis outburst behavior. We conclude from both vertical structure models (§5.) and time dependent diffusion models (§6.) that there is a maximum attainable steady state mass flux: $\dot{M}_{crit} = 5 \times 10^{-7}$ M$_\odot$/ yr: a value essentially independent of the form or magnitude of $\alpha$, above which protostellar disks are subject to a self–regulated, global thermal instability resulting in repetitive FU Orionis type events. Using the time scales associated with such outbursts we estimate local $\alpha$ values between $10^{-3}$ and $10^{-4}$. Results and limitations of this model are summarized in §7.; consequences are discussed in §8..

## 2. BACKGROUND

### 2.1. Observational history

In 1936 prototype FU Orionis flared six magnitudes in the short span of a year (Herbig 1966). In the almost sixty years since, the brightness of FU Ori has remained essentially constant. Herbig's earliest proposal attributes the outburst to a dramatic once per star gravitational collapse marking the appearance of the central protostellar object onto the convective portion of its Hayashi track. Subsequent discoveries of FU Ori type variables ("Fuors") V1057 Cyg and V1515 Cyg made the once per star hypothesis untenable, and Herbig (1977) instead proposed on statistical grounds that the FU Orionis phenomenon was a recurrent event which happened to all T Tauri systems approximately every ten thousand years. That the phenomenon is associated with low mass YSOs is supported by the one pre-outburst spectrum of V1057 Cyg showing typical T Tauri features.

The existence of additional FU Orionis outbursts have been derived from historical records, bringing the total number of confirmed outbursts to six. Properties of these events (many of the time scales are merely constraints) are summarized in Table 1. On the basis of spectral features such as those described below, five additional objects have been proposed as members of the class in outburst including binaries RNO 1B/1C, in which both members are apparently in outburst and Z CMa (the binarity of this object may account for its large derived luminosity). In the discussion which follows we note that all statistical results are based on these small numbers of observed events.

## 2.2. Spectral characteristics

While all T Tauri systems are highly variable, FU Orionis outbursts are a particularly powerful form of outburst accompanied by distinctive spectral changes not seen in lower level T Tauri activity. T Tauri spectra are typically dominated by single temperature M or K type (presumably stellar) black bodies (3000 – 4000K) with strong H$\alpha$, Ca II, and CO emission lines and significant infrared and ultraviolet excesses (IRE and UVE) such that $L_* \approx L_{IRE} \approx L_{UVE} \approx 1$ L$_\odot$. FU Orionis systems, in contrast, have spectra generally dominated by F or G type spectra (6000 – 7000K) with low gravity, supergiant absorption features in the optical. Assigned spectral types are later at longer wavelengths. FU Ori spectra have no obvious single black body source: $L_* \ll L_{IRE} \approx$ several 100 L$_\odot$, and no observed UV excess (Kenyon et al. 1989; the lack of UV excess has been attributed to a radial distribution of energy in the thicker Fuor disks [Bell, Lin, & Papaloizou 1991, Popham et al. 1993]). Spectra in all but the earliest phases of outburst are self-consistently reproduced by a simple constant mass flux accretion disk model (Kenyon, Hartmann, & Hewett 1988; Kenyon & Hartmann 1991).

In Fuors lines which are typically in emission in T Tauri systems (such as Balmer and CO) are generally in absorption (Carr 1988). H$\alpha$ displays a P Cygni profile with weak emission and deep blue–shifted absorption often extending several hundred km/sec from the line center. In all small scale variability, including "Exor" events (after prototype EX Lupus; Herbig 1989) characterized by rather large (2–5 magnitude) but short–lived (months to a few years) outbursts, T Tauri type emission lines are retained. That H$\alpha$ emission persisted in Fuor V1057 Cyg for several years after peak light (cf. Figure 6 in Herbig 1977) before being superseded by the typical Fuor H$\alpha$ profile gives weight to Herbig's suggestion (1989) that the same physical mechanism may be responsible for both Exor and Fuor outbursts. Nevertheless, we limit our modeling in this work to fits of the more dramatic and less frequent FU Orionis type outbursts.

## 2.3. The outburst as an accretion event

During FU Orionis outbursts, it is observed that at longer wavelengths one sees (1) a systematically cooler spectral type (Herbig 1977) and (2) a decreased separation of doubling in absorption lines (Hartmann and Kenyon 1987a, b; Welty, et al. 1990, 1992). These observations may be used as evidence that the energy released during outburst is provided by a self–luminous accretion disk for which matter emitting at successively longer wavelengths is both cooler and has a slower rotation velocity. In accretion disks, for which $L_{acc} \sim GM_*\dot{M}/R_*$, an optical brightening of 4–6 magnitudes corresponds to an increased mass flux factor such that $\dot{M}_{FU}$ should be $(40 - 250) \times \dot{M}_{TT}$ ($\dot{M}_{TT} = (1 - 30) \times 10^{-8}$ M$_\odot$/yr). While the case for self–luminous disks in T Tauri objects is largely circumstantial, dereddened broad band spectral points are well fit by the assumption that **all** energy in Fuor systems comes from a disk with an accretion rate $\dot{M}_{FU}$ between $10^{-5}$ and $10^{-3}$ M$_\odot$/yr (Table 1) (Hartmann & Kenyon 1985, 1987a, b; Kenyon et al. 1988; Kenyon & Hartmann 1991).

## 2.4. Observational constraints

Observations of YSO systems provide useful constraints for outburst models. Quiescent T Tauri disks have surface temperatures up to 3000K. Mass fluxes ($\dot{M}_{TT}$) derived from models of UV excesses as boundary layer emission range up to several $10^{-7}$ M$_\odot$/yr (Bertout et al. 1988). Outbursts are between 4 and 6 magnitudes in the optical, generally less in the infrared (Simon & Joyce 1988; Kenyon & Hartmann 1991). Spectral energy distributions of Fuors FU Ori and V1057 Cyg are fit to steady mass flux disks with maximum disk surface temperatures between 6000 and 7200K and mass fluxes ($\dot{M}_{FU}$) of $(5 - 40) \times 10^{-5}$ M$_\odot$/yr (Kenyon et al. 1988, taking M$_* = 1$M$_\odot$). Lower and higher mass fluxes have been indicated for other Fuors (Table 1); we take our constraint on $\dot{M}_{FU}$ to be $(1 - 100) \times 10^{-5}$ M$_\odot$/yr.

The outburst time scale: $\tau_{rise}$, is based upon the three Fuors: FU Ori, V1057 Cyg, and V1515 Cyg, with time resolved photometry during outburst and varies from 1 to 20 years (cf. Table 1). Time spent in the high state: $\tau_{high}$, is a lower limit as no Fuor has been seen to drop back to a normal T Tauri state. Prototype FU Orionis has shown only minimal fading in the 60 years since outburst while V1057 Cyg initially faded by several magnitudes over a ten year period following outburst before apparently stabilizing some three magnitudes brighter than its pre-outburst value. Postulated class member Z CMa seems to have been near its current brightness 130 years ago (Hartmann et al. 1989). We take $\tau_{high}$ to be between 80 and 150 years as a constraint to our models.

The interval between Fuor events, $\tau_{FU}$, in a given object is only statistically constrained and depends critically upon our assumptions as to which systems we believe are subject to outburst. We start with the simplest assumption in which all low mass YSOs go through such events with equal probability. Six Fuor events have been observed since the turn of the century out of a population of 750 catalogued YSOs (e.g., Herbig & Bell 1988) suggesting $\approx 10^{-4}$ events per year for each low mass YSO or 10,000 years between events. Because some events surely have been missed and also because we suspect that only some fraction of observed T Tauri stars are subject to such events (§8.) we take the time between outbursts, $\tau_{FU}$, to be on the order of 1,000 years. Partial confirmation of this value is given by the 1,000 year spacing of Herbig–Haro objects (shock heated "bullets" hypothesized to be the result of Fuor events [Dopita 1978; Reipurth 1985; and Hartmann & Kenyon 1985]) in the outflow of postulated Fuor L1551 (Stocke et al. 1988). The time scale $\tau_{FU}$ is the least accurately determined of the time scales we will consider.

## 3. DISK STABILITY

### 3.1. Time scales of viscous disks

With observed surface temperatures between 6000 and 7000K, the midplanes of the inner regions of optically thick Fuor disks are sufficiently hot for the ionization of hydrogen ($T_c \gtrsim 10,000$K). Theoretical models which investigate the idea that FU Orionis events are primarily the result of a global thermal instability in the protostellar accretion disk were explored by Lin & Papaloizou (1985) and later developed by Clarke, Lin, & Papaloizou (1989) and Clarke, Lin, & Pringle (1990). In the protostellar case as in the CV case, the thermal instability arises for sufficiently high disk mass fluxes such that hydrogen is expected to be ionized at some intermediate radius. Due to the rapidly increasing $H^-$ opacity near the ionization of hydrogen, regions of partial ionization are thermally unstable and will heat (or cool) until a new opacity regime is reached in which the region is fully ionized (or neutral). In this section we discuss the time scales associated with thermal instability.

According to the *ad hoc* $\alpha$ viscosity prescription $\nu = \alpha c_s H$, where $c_s = \sqrt{R_g T/\mu}$ is the isothermal sound speed ($R_g$ is the gas constant and $\mu$ the mean molecular weight) and $H(\approx c_s/\Omega)$ is the local pressure scale height. Upon entering the unstable regime caused by the initiation of hydrogen ionization, the thermal instability grows until the unstable annulus becomes fully ionized on the local thermal time scale given by energy content divided by energy generation:

$$\tau_{therm} \simeq \frac{C_v \Sigma T_c}{\frac{9}{4}\Sigma \nu \Omega^2} \approx \frac{H^2}{\nu} \approx \frac{1}{\alpha \Omega},$$

where $\Sigma$ is the full plane surface density, $\Omega = \Omega_K = \sqrt{GM_*/r^3}$ is the Keplerian period, and $C_v$ is the specific heat at constant volume. As the temperature increases, the effective local mass flux also increases, and the hot regions expand and begin to propagate radially through the disk. The ionization front which separates hot and cold regions propagates to a particular radius on the time scale determined by the maximum speed allowed by the Rayleigh stability criterion for axisymmetric flows (Lin et al. 1985):

$$\tau_{prop} \simeq \frac{r}{\alpha c_s} \approx \frac{Hr}{\nu} \approx \left(\frac{r}{H}\right) \tau_{therm}.$$

During outburst, the hot, ionized, inner regions deplete on the viscous diffusion time scale,

$$\tau_{visc} \simeq \frac{r^2}{\nu} \approx \left(\frac{r}{H}\right)^2 \tau_{therm}.$$

When the disk cools below the ionization temperature of hydrogen, the thermal front retreats inward on the propagation time scale leaving the disk once again fully neutral. If, as in typical YSO disks, $H/r \approx 0.1$ (Lin & Papaloizou 1985), the ratio $\tau_{prop}/\tau_{visc} \approx H/r$ is suggestively close to the ratio: $\tau_{rise}/\tau_{high}$, thought to occur in Fuor systems. Between outbursts, matter builds up until the disk once again becomes sufficiently hot to ionize hydrogen. This cycle results in self–regulated, periodic outbursts.

## 3.2. Previous models of YSO outbursts

Clarke et al. (1989) investigate the propagation of thermal fronts in the context of protostellar accretion disks using a one dimensional radial time dependent diffusion code. They find that the propagation of a thermal front may be quenched due to the radial advection of heat in the presence of a large radial temperature gradient if the local pressure scale height is large. Since the thermal instability always occurs at the temperature of the ionization of hydrogen: $T_c \approx 10^4$ K, the scale height near the ionization front $H \approx c_s/\Omega \sim r^{3/2}$, and so the disk thickness increases rapidly with $r$ as the front propagates outward. The front propagates outward until it is sufficiently thick to be stabilized by nonlocal effects. Clarke et al. (1989) show that disks fed the high mass flux of $10^{-4}$ M$_\odot$/ yr attain a steady state with the thermal front stalled at about half an AU. Disks fed the slightly lower mass flux of $10^{-5}$ M$_\odot$/ yr do not provide scale heights sufficient to stabilize the front and are found to be subject to radial excursions of the ionization front on time scales of a few years. We note that this lower limit for theoretical quenching of front propagation of a few times $10^{-5}$ M$_\odot$/ yr is coincident with the estimated lower limit of Fuors mass fluxes as expressed in Table 1.

In a subsequent paper, Clarke et al. (1990) show that the observed time scales and magnitudes of FU Orionis outbursts may be reproduced by applying a smoothed top hat surface density perturbation to a low mass flux T Tauri disk. Using the large perturbation of $\Delta\Sigma/\Sigma = 50$ placed a quarter of an AU from the central object, Clarke et al. were able to recreate FU Orionis time scales finding $\tau_{rise} \approx 8$ yrs and $\tau_{high} \approx 100$ yrs. These results were obtained with a value of $\alpha = 10^{-3}$. Further investigation suggests that such a large perturbation if caused by a stellar encounter would be likely to destroy the disk (Clarke & Pringle 1991) requiring a reformation of the disk for each FU Orionis outburst (Clarke et al. 1990).

## 4. METHOD

We propose the scenario in which FU Orionis outbursts occur in a protostellar accretion disk due to the self–regulating effects of the thermal instability without the need for external perturbation. We postulate that a disk subject to Fuor outbursts receives mass (perhaps from a remnant infalling envelope) with mass flux, $\dot{M}_{in}$, between the T Tauri value of $\dot{M}_{TT}$ = $(1 - 30) \times 10^{-8}$ $M_{\odot}/$ yr and the FU Orionis value of $\dot{M}_{FU} = (1 - 30) \times 10^{-5}$ $M_{\odot}/$ yr. An accretion disk is self–regulated by the thermal instability in such a way that it is in the low mass flux T Tauri state with a neutral disk for most of its pre- Main Sequence lifetime. Surface densities in the disk build up during the low state until the inner annuli become sufficiently hot to ionize hydrogen and an FU Orionis outburst begins. An ionization front propagates outward from the ignition point on a propagation time scale (cf. §3.1.) until being stabilized at some radius, $R_{limit}$; during the outburst mass falls onto the central object at the rate $\dot{M}_{acc} \approx \dot{M}_{FU}$ $(>> \dot{M}_{in})$. In regions beyond the radial propagation of the front $(r > R_{limit})$ mass flux remains steady and is equal at all times to $\dot{M}_{in}$. Enhanced mass flux in the inner ionized region continues until the surface density drops (on a viscous time scale) below its lowest sustainable high state value. The ionization front then retreats inward and the system drops down to a neutral state during which $\dot{M}_{acc} \approx \dot{M}_{TT}$ $(<< \dot{M}_{in})$ leaving a greatly depleted inner region into which mass begins accumulating for the next outburst.

We use two models to self–consistently describe conditions in the protostellar accretion disk. (1) One dimensional, hydrostatic, vertical structure (VS) models allow high resolution of physical conditions (including convective stability) from disk midplane to photospheric optical depths. From these models, we derive the local radiative loss in the vertical direction as a function of $\Sigma$ (full plane surface density) and $T_c$ (central temperature). (2) Vertically averaged, radial, time dependent diffusion (TDD) models are used with only input mass flux $\dot{M}_{in}$ and viscous efficiency $\alpha$ as parameters in FU Orionis light curve fitting. In both models $M_*$ is assumed equal to 1 $M_\odot$. Both models also assume axial symmetry and use the same opacity law and detailed equation of state. The analytic opacity law accounts for the most recent Alexander et al. (1989) opacities and is described in Appendix A.. The equation of state, for which we are indebted to Dr. P. H. Bodenheimer, assumes solar composition, and includes the effects of ionized and molecular states of hydrogen as well as singly and doubly ionized states of helium.

## 4.1. Vertical structure models

Vertical structure models have two purposes in this work. Calculations are first done assuming vertical thermal balance (VTB) in order (a) to examine local stability for a range of conditions including variation of the parameter $\alpha$ and the effect of the presence or absence of convection and (b) to make some estimates about the radial propagation of outbursts. Second, detailed *non*–VTB vertical structure models are used to estimate vertical radiative losses for the time dependent diffusion models. In both cases we assume hydrostatic equilibrium for the disk because the time scale required to attain such a state is the most rapid time scale in the disk, the dynamical time scale ($\tau_{dyn} \approx 1/\Omega$). In the VTB models, we make use of the standard thin disk approximation (e.g., Lin & Papaloizou 1985) in which local vertical energy losses:

$$Q^- \equiv 2F_s = 2\sigma T_{eff}^4,$$

where $F_s$ is the surface flux, are exactly balanced by local viscous energy generation:

$$Q^+ \equiv \int\limits_{-H_d}^{H_d} \frac{9}{4}\rho\nu\Omega^2 \mathrm{d}z \approx \frac{9}{4}\Sigma\nu_c\Omega^2,$$

where $\nu_c$ is the vertically averaged viscosity (often taken to be equal to the midplane viscosity). In the non–VTB vertical structure models, we introduce a technique which allows for the systematic variation of the relationship between $Q^-$ and $Q^+$. We are required to account for deviations from $Q^- = Q^+$, because in the time dependent diffusion models conditions are expected to deviate strongly from vertical thermal balance and nonlocal energy sources such as the advective heat transport and radial radiative diffusion are likely to contribute significantly to the internal energy flow.

### 4.1.1. Vertical thermal balance models

First we discuss the standard vertical thermal balance VS models. When $Q^- = Q^+$, the local surface flux, $F_s$ (or equivalently the effective temperature, $T_{eff}$), of a given annulus is uniquely determined from the specification of the local magnitude of the mass flux, $\dot{M}$. (Although there is some evidence that radiative feedback may be a factor in the long term stability of accretion disks [Bell, Lin, & Ruden 1991], throughout this work we neglect effects due to the disk's reprocessing of light from the central object.) In a constant mass flux disk, the vertically averaged $\phi$ component of the angular momentum equation becomes (e.g., Pringle 1981)

$$\Sigma \nu_c = \frac{\dot{M}}{3\pi} \left( 1 - \beta \sqrt{\frac{R_*}{r}} \right),\tag{1}$$

where $\beta \equiv \Omega_{\rm disk}(R_*)/\Omega_{\rm K}(R_*)$. This equation is combined with the VTB assumption $Q^- = Q^+$ to obtain a relation between the surface flux of a given annulus and its mass flux:

$$\sigma T_{eff}^4 = \frac{3 {\rm GM}_* \dot{M}}{8\pi r^3} \left( 1 - \beta \sqrt{\frac{R_*}{r}} \right).\tag{2}$$

In the VTB vertical structure models, we take $\beta = 1$ (the "no torque" boundary condition which assumes $\Omega_* << \Omega_{\rm K}(R_*)$) consistent with a narrow boundary layer (e.g., Pringle 1981), and ${\rm R}_* = 3~{\rm R}_\odot$, a typical T Tauri radius. Differentiation with respect to $r$ of Eq. (2) gives an expected peak surface flux at $\left( \frac{7}{6}\beta \right)^2 {\rm R}_*$ which for our parameters occurs at $4.1~{\rm R}_\odot$; we will use this radius as a characteristic inner radius when examining equilibrium curves in §5.. For vertical integrations to be used in the time dependent radial routine, we take $\beta = 0$ (which assumes Keplerian rotation continues in to ${\rm R}_*$) to allow calculation of $F_s$ all the way to the inner edge of the disk where $r = {\rm R}_*$.

A vertical structure model is calculated for a particular $\alpha$ given a mass flux, $\dot{M}$, and radius, $r$. Integration begins by assuming a value of the photospheric half thickness $H_d$ (*not* in general equal to the local pressure scale height, $H \approx c_s/\Omega$). The surface pressure, $P_s$, and density, $\rho_s$, are found iteratively by requiring that surface pressure be equal to the weight of the atmosphere above it with the assumption that $\tau_s \equiv \int\limits_{H_d}^{\infty} \kappa \rho {\rm d}z \approx \kappa_s \int\limits_{H_d}^{\infty} \rho {\rm d}z$ where $\kappa_s = \kappa(T_{eff}, \rho_s)$:

$$P_s = g \int\limits_{H_d}^{\infty} \rho {\rm d}z \approx g \frac{\tau_s}{\kappa_s}.$$

We start at a surface optical depth $\tau_s = 2/3$ (thus assuming that the disk is optically thick in the vertical direction) and using an explicit, finite difference method, integrate inward to the midplane. Pressure and flux are staggered half a zone from temperature and density for computational accuracy.

The condition of hydrostatic equilibrium is imposed *via* the $z$ component of the equation of motion:

$$\frac{\partial P}{\partial z} = -\rho g = \frac{-\rho \mathrm{GM}_* z}{(r^2 + z^2)^{\frac{3}{2}}}.$$

We *do not* make the conventional assumption $g = \Omega^2 z$ and thus do not strictly require $H \ll r$. Flux and temperature are integrated with the following conditions: (1) Energy is assumed to be generated through local viscous stress i.e., $\frac{\partial F}{\partial z} = \frac{9}{4}\rho\nu\Omega^2$. We use the local alpha law viscosity where $\nu = \alpha c_s^2/\Omega$; the form of $\alpha$ is discussed below. (2) Both radiative and convective energy transport are included. For convection we use the simple mixing length theory (MLT: Cox & Giuli 1968; Lin & Papaloizou 1985 with Eq. 73 corrected) in which mixing occurs over the local pressure scale height $\Lambda = \min(H, H_d)$ (thus taking the MLT $\alpha$ parameter to be 1.0). Despite its approximate nature, MLT used in one dimensional models reproduces quite well nebular structure found from detailed two dimensional hydrodynamic calculations (Różyczka, Bodenheimer, & Bell 1994; Kley, Papaloizou, & Lin 1993). (3) We allow propagation of energy in optically thin zones at no faster than the speed of light: diffusion is flux limited as in Bodenheimer et al. (1990). Upon reaching the midplane, the disk half thickness, $H_d$, is successively modified to fulfill the symmetry requirement that the midplane flux be zero. Convective calculations use 100 zones and radiative calculations use 500 zones. The grid spacing is a self–adapting mesh designed to resolve rapid gradient changes and superadiabatic regions.

Equilibrium curves are created at given radius from a series of VS models for which the free parameter $\dot{\mathrm{M}}$ is varied over the range of interest. Curves calculated with these models are highly dependent on the value of $\alpha$ and on the form of the opacity: results are presented in §5..

### 4.1.2. Vertical models not in thermal balance

We now discuss modifications to the above method required to account for departures from vertical thermal balance. The VTB procedure discussed above produces results with only one free parameter: $\dot{\mathrm{M}}$. With the assumption that $Q^- = Q^+$, varying $\dot{\mathrm{M}}$ (while keeping $\alpha$ and $r$ constant), creates a simple relationship between $\Sigma$ and $T_c$; any arbitrary combination of $\Sigma$ and $T_c$ (such as might arise in the TDD models) may not necessarily fall on the line defined by $Q^- = Q^+$. In fact we *expect* strong deviations from VTB to occur during outburst. Our technique allows calculation of detailed VS models while self–consistently accounting for all radial and time dependent terms in the energy equation (5) below.

We introduce a second degree of freedom into the VS models with the definition of the fractional heat imbalance $\mathcal{F}$ which allows us account for departures from VTB in a systematic way:

$$\mathcal{F} \equiv \frac{Q^- - Q^+}{Q^- + Q^+}.$$

The effects of this term are distributed evenly through the vertical thickness of the disk by the inclusion of a factor in the flux calculation such that $\frac{\Delta F}{\Delta z} = \frac{9}{4}\rho(z)\nu(z)\Omega^2 + Q^- \frac{2\mathcal{F}}{(1+\mathcal{F})} H_d^{-1}$ which when integrated over $z$, reduces to $F_s = Q^-$. Although the parameter $\mathcal{F}$ is not explicitly used in the TDD models, its inclusion in the VS models allows us to self–consistently calculate $F_s(\Sigma, T_c)$ for all possible combinations of $\Sigma$ and $T_c$. One of the effects of the implementation of this new technique is that in some TDD situations even if $Q^- \neq Q^+$, thermal equilibrium may still be maintained ($dT/dt = 0$) provided the energy flux due to nonlocal effects is balanced by $\mathcal{F} \times (Q^- + Q^+)$.

Non–VTB models are calculated for a range of radii, temperatures and degrees of departure from thermal equilibrium. Results are organized into tables which for each radius and $\alpha$ give $F_s(\Sigma, T_c)$, or equivalently $T_{eff}(\Sigma, T_c)$, for use in the time dependent models. As an example, we display a surface plot of $\Sigma(T_c, \mathcal{F})$ at a radius of 20.3 $R_\odot$ ($\approx 0.1$ AU) for $\alpha = 10^{-4}$ in Figure 1. The line at the bottom traces $\Sigma(T_c)$ for the VTB case: $\mathcal{F}=0$. The disk is unstable in the region between the arrows where surface density decreases for increasing central temperature. An analogous grid of $T_{eff}(T_c, \mathcal{F})$ is also produced but not shown. In the TDD models, the unique $T_c$ and $\Sigma$ pair are located in the $\Sigma(T_c, \mathcal{F})$ grid and the analogous value is read off of the $T_{eff}(T_c, \mathcal{F})$ grid. We find that $T_{eff}$ varies by 30% from $\mathcal{F}=0$ to $\mathcal{F}= \pm 0.9$ which confirms the necessity of using this technique in systems which are expected to be far from thermal equilibrium.

### 4.2. Radial time dependent diffusion models

Radial time dependent diffusion (TDD) models are calculated with 72 logarithmically spaced Eulerian grid points with boundaries $R_{inner} = 3$ $R_\odot$ and $R_{outer} = 100$ $R_\odot \approx 0.5$ AU. We examine a range of input mass fluxes: $\dot{M}_{in} = 10^{-7}$ to $10^{-5}$ $M_\odot/yr$, and a range of $\alpha$'s: $10^{-1}$ to $10^{-4}$. Time steps are diffusion limited in the early settling stages of the calculation, limited by the smallest zone sound crossing time when approaching steady state, and limited by the photon zone–crossing time in a few trial cases; because of the explicit nature of the radial flux calculation, this much shorter time step is technically required, but results presented in § 6.2.4. demonstrate that the essential results of the model are not affected by our larger time step. At each step we calculate values of full plane surface density, $\Sigma$, and central temperature, $T (\equiv T_c)$, for each radius from the simultaneous solution of the equation of continuity:

$$\frac{\partial \Sigma}{\partial t} = \frac{1}{r}\frac{\partial}{\partial r}\left(\Sigma U_r r\right) + S_{\Sigma}(r), \tag{3}$$

the $\phi$ component of the equation of motion:

$$\Sigma U_r r = \frac{\dot{M}(r)}{2\pi} = -3r^{\frac{1}{2}}\frac{\partial}{\partial r}\Sigma \nu r^{\frac{1}{2}}, \tag{4}$$

and the time dependent energy equation:

$$
\begin{aligned}
C_v \Sigma \frac{\partial T}{\partial t} &= \frac{9}{4}\Sigma\nu\Omega^2 - 2\sigma T_{eff}^4 - 2\left(\frac{H}{r}\right)\frac{\partial}{\partial r}\left\{r\left(\frac{ac}{3\kappa_c\rho_c}\frac{\partial T^4}{\partial r}\right)\right\} \\
&+ C_v\left[U_r\Sigma\frac{\partial T}{\partial r} + (\Gamma_3 - 1)T\left\{\frac{\partial \Sigma}{\partial t} - U_r\frac{\partial \Sigma}{\partial r}\right\}\right].
\end{aligned}
\tag{5}
$$

All equations are vertically averaged and assume that rotation is Keplerian on cylinders. In Eq. (3), the source term $S_{\Sigma}(r)$ simulates infall which is added at a constant rate in a Gaussian distribution over several zones well beyond the thermally unstable region[3] such that $\int_{R_1}^{R_2} S_{\Sigma}(r)\mathrm{d}r = \dot{M}_{in}$ and is added at the ambient temperature of the disk. In the energy equation (5), the specific heat, $C_v$, and third adiabatic exponent, $\Gamma_3$, include the effects of ionization and recombination in a solar composition mixture. The first term of the energy equation describes the local energy generation due to viscous dissipation ($Q^+$). The second term accounts for energy lost by radiation from the two surfaces of the disk ($Q^-$): effective temperature is found from the results of the detailed non–VTB vertical structure models as described above. The term with ($H/r$) describes contribution from radiative diffusion (divergence of the flux), the $\partial T/\partial r$ term describes advection (energy moves along with mass), and the final two terms describe the PdV work.

The local pressure scale height, $H$, is found from the relation $H = \sqrt{\left\{\frac{R_g T}{\mu\Omega^2} + \left(\frac{2a T^4}{\Omega^2\Sigma}\right)^2\right\}}$ which is the usual expression: $H = c_s/\Omega$, with a correction to account for the effects of radiation pressure. Local midplane density: $\rho_c$, is given by $\Sigma/(2H)$ and opacity by $\kappa_c \equiv \kappa(T, \rho_c)$. Viscosity is derived self–consistently from the detailed VS models such that $\nu = \frac{1}{\Sigma}\int_{-H_d}^{H_d}\rho\nu\mathrm{d}z$. At the inner boundary $\Sigma_0 = 0$: a condition in which mass accretes freely onto the central object; $\dot{M}_{acc} \equiv \dot{M}(R_{inner})$. At the outer boundary mass flux is set equal to zero to prevent the outward spread of the input mass flux which would lead to a gradual depletion of the disk. For several trial cases we alter the inner boundary condition to one with a zero radial gradient in mass flux for which $(\Sigma\nu)_0 = \Sigma_1\nu_1$; the change of boundary condition is found to have no impact on the resultant time scales. Models are followed until the integrated disk mass stabilizes to ensure the erasure of arbitrary initial conditions.

---

[3]This is *unlike* the CV case in which mass must be added inside the thermally unstable region and results are therefore sensitively dependent on the details of the source term.

### 4.3. The importance of self–gravity

In our analysis of disk evolution, we have neglected the effects due to self–gravity of the disk. This approximation is adequate provided the gravitational stability parameter (Safronov 1960; Toomre 1964) $Q \equiv c_s\Omega/\pi G\Sigma$ is much larger than unity. An approximate form for $\Sigma(r)$ in a steady state disk can be derived from Eq. (1) by assuming a viscosity. Using an $\alpha$ law viscosity and the approximation $H \approx c_s/\Omega$ and taking $H/r \approx$ constant, one can show that $\Sigma \sim r^{-1/2}$, and that $Q_T \sim r^{-3/2}$. Further one can show that the radius for which $Q_T = 1$ is given by

$$R(Q_T = 1) = 1.5 AU \left\{ \left(\frac{\alpha}{10^{-3}}\right)^{2/3} \left(\frac{H/r}{0.1}\right)^2 \left(\frac{\dot{M}_{in}}{10^{-5} \ M_\odot/\, yr}\right)^{-2/3} \left(\frac{M_*}{1M_\odot}\right) \right\}.$$

It is also interesting to estimate the mass of the disk interior to this point:

$$M_{disk}(Q_T = 1) = \int\limits_{R_*}^{R(Q_T=1)} \Sigma 2\pi r dr = \frac{4}{3}\frac{H}{r}M_* \approx 0.133 M_*.$$

Although $R(Q_T = 1)$ depends upon the unknowns $\alpha$ and $\dot{M}_{in}$, $M_{disk}(Q_T = 1)$ does not. This analysis indicates that in the thermally unstable regions of the disk ($r < 1/4$ AU), the effect of self–gravity is negligible and therefore it will not strongly modify the thermal instability process. In addition, we monitor the disk mass during our calculation to ensure that it does not rise above 0.133 $M_*$ thereby ensuring that self–gravity is negligible throughout the radial extent ($3\ R_\odot < r < 1/2$ AU) of our models.

## 5. VERTICAL STRUCTURE RESULTS

### 5.1. Equilibrium curves

We show in Figure 2 the results of vertical structure calculations of equilibrium curves ($Q^- = Q^+$) at the characteristic inner disk radius of 4.1 $R_\odot$ (cf. §4.1.1.). The four solid curves in each panel represent disks with different values of $\alpha$: $10^{-1}$, $10^{-2}$, $10^{-3}$, and $10^{-4}$.

We define $\Sigma_A$ to be the highest stable low state surface density for a given $\alpha$; $\Sigma_A$ for $\alpha$ = $10^{-4}$ is marked by the vertical arrow in Figure 2a. If at any radius the surface density increases above $\Sigma_A$, the disk can no longer stay stably on the lower branch and strong local heating begins. We likewise define $\Sigma_B (< \Sigma_A)$ to be the lowest stable high state surface density for a given $\alpha$. Any ionized annulus which drops below $\Sigma_B$ will recombine and cool on a thermal time scale. The largest stable mass flux at any radius is the mass flux associated with $\Sigma_A$; this value: $\dot{M}(\Sigma_A)$, is indicated by the horizontal arrow in Figure 2a for $\alpha = 10^{-4}$. It can be seen that both the largest stable mass flux and its associated effective temperature are essentially independent of $\alpha$. We define $\dot{M}_{crit} = 5 \times 10^{-7}$ $M_\odot / \mathrm{yr}$ to be the value of $\dot{M}(\Sigma_A)$ at the characteristic radius *independent of the value of* $\alpha$. This value is in agreement with observations of the boundary layers of the highest mass flux T Tauri objects.

As expected from Eq. (1), it can be seen in Figure 2a that lowering $\alpha$ raises the local surface density by approximately a factor of $\alpha$. Midplane temperatures are also strongly dependent on the magnitude of $\alpha$. This figure demonstrates not only the necessity of using time dependent models to constrain protostellar disk surface densities and midplane temperatures but also just how many orders of magnitude of uncertainty exist. It can also be seen from Figure 2d that the vertical thickness of the disk increases greatly as a result of the ionization of hydrogen at large mass fluxes. We expect radial effects to become significant during outburst.

In Figure 2, radiative VS models (detailed VS models calculated with no convective energy transport) are shown in each panel as dashed lines for the highest and lowest values of $\alpha$. For the highest value of $\alpha$ ($10^{-1}$), the radiative and convective equilibrium curves are nearly identical (see also Faulkner et al. 1983), but for the lowest value of $\alpha$ ($10^{-4}$), the transition region is quite different for convective and radiative models. Not only would one estimate a critical mass flux that would be too low by looking only at the radiative models (and one that was highly dependent on the value of $\alpha$), but one would also predict a much lower effective temperature and consequently seriously underestimate the radiative cooling near the critical transition region. Away from the transition region, however, for both upper and lower branches the radiative approximation is quite good.

### 5.2. Estimates of radial propagation

We show in Figure 3, the radial dependence of equilibrium (VTB) curves: $T_{eff}(\Sigma)$ and $\dot{M}(\Sigma)$ for $r = 5$, 10, 20, 40, and 80 $R_\odot$. These models are calculated with our standard viscosity for which $\alpha_c = 10^{-4}$ and $\alpha_h = 10^{-3}$ (this form is described in the next section).

While the critical effective temperature ($\approx$ 2000K) is only weakly dependent on radius, critical surface densities ($\Sigma_A$) and largest stable mass fluxes ($\dot{M}(\Sigma_A)$) are strong functions of radius. Recalling from the previous section that the largest stable mass flux is essentially independent of $\alpha$, we use results of the vertical structure models to estimate the radial propagation of the thermal front as a function of the input mass flux.

For our conditions where $R_* = 3$ $R_\odot$ and $M_* = 1$ $M_\odot$, for $\dot{M}_{in}$ less than $\dot{M}_{crit}$ = $5 \times 10^{-7}$ $M_\odot$/ yr the entire disk can remain stably in the low mass flux T Tauri state indefinitely. This is possible because for low input mass fluxes (such as the $10^{-7}$ $M_\odot$/ yr indicated by the lowest arrow in Figure 3b), all annuli have accessible lower branch surface densities; that is to say, in this low mass flux state, for all annuli there exist surface densities for which $\dot{M}_{in}$ can be transported such that that hydrogen does not ionize. For input mass fluxes above $\dot{M}_{crit}$ (such as the $3 \times 10^{-6}$ $M_\odot$/ yr indicated by the middle arrow in Figure 3b) this is not the case.

We examine a disk transporting the representative input mass flux rate of $\dot{M}_{in}$ = $3 \times 10^{-6}$ $M_\odot$/ yr which corresponds to the middle arrow in Fig. 3b. The curves representing 40 and 80 $R_\odot$ annuli for this disk have accessible stable points on the lower branch. At 5 and 10 $R_\odot$, however, the surface densities corresponding to the input mass flux are on backwards sloping parts of the equilibrium curves and are therefore expected to be unstable on the local thermal time scale. The 20 $R_\odot$ case appears to be marginal. For $\dot{M}_{in} = 3 \times 10^{-6}$ $M_\odot$/ yr, radii inside about 20 $R_\odot$ *must* undergo thermal outburst in order to transport (in a time averaged way) the mass flux $\dot{M}_{in}$ being supplied from the outer disk. For this particular mass flux we would conclude that the ionization front travels out to some 20 $R_\odot$. For larger $\dot{M}_{in}$, the limiting radius ($R_{limit}$) is correspondingly larger.

The magnitude of $R_{limit}$ can be quantified. By making use of Eq. (2), by assuming $R_{limit} >> R_*$, and by taking $T_{eff} = $ 2000K as a typical critical effective temperature (Fig. 3a), we can estimate *independent of* $\alpha$ the dependence of $R_{limit}$ on $\dot{M}_{in}$ and $M_*$:

$$R_{limit} = 20R_\odot \left( \frac{\dot{M}_{in}}{3 \times 10^{-6} \text{ M}_\odot/\text{yr}} \right)^{\frac{1}{3}} \left( \frac{M_*}{M_\odot} \right)^{\frac{1}{3}} \left( \frac{T_{eff}}{2000} \right)^{-\frac{4}{3}}.$$

In practice, the thermal fronts over run this radius to some degree; this provides an estimate of the *minimum* radial propagation of the front.

### 5.3. Functional form of $\alpha$

For most time dependent models we increase the value of $\alpha$ in moving from the low state to the high state following the practice common in CV outburst modeling (e.g., Smak 1983; Mineshige & Osaki 1985; Cannizzo 1993). We choose to smooth the transition by following the ionization fraction of hydrogen ($\chi_i$) such that $\alpha = \alpha_c + (\alpha_h - \alpha_c)\chi_i^{1/4}$ where $\alpha_c$ is the value of $\alpha$ in the cool, neutral state and $\alpha_h$ is the value of $\alpha$ in the hot, ionized state.

According to the arguments made in the previous section, the radial propagation of the front depends strongly on $\dot{M}_{in}$ but is nearly independent of $\alpha$. The time spent in outburst is primarily determined by the viscous time scale in the hot state and thus by the magnitude of $\alpha_h$ and is essentially independent of $\alpha_c$. Using the time scale arguments in §3.1., we derive the low value of $\alpha_h \approx 3 \times 10^{-4}$ suggested by requiring 100 years for a front to viscously deplete the disk out to 20 $R_\odot$ (we take $c_s$ to be the isothermal sound speed in a fully ionized medium at $10^5$ K). This estimate agrees with the value needed by Clarke et al. (1990) to obtain observed FU Orionis outburst time scales.

The time between outbursts depends upon the contrast between $\alpha_h$ and $\alpha_c$: for a given $\alpha_c$, a larger $\alpha_h$ results in a longer time between outbursts. Increasing the ratio of $\alpha_h$ to $\alpha_c$ has the effect of increasing the contrast between the critical surface densities: $\Sigma_A$ and $\Sigma_B$. Immediately after outburst, the surface density of the inner disk is left in a state where the regions subject to outburst are such that $\Sigma(r) \approx \Sigma_B(r)$. The time spent in the low state is then determined by viscous processes such that $\delta t \propto \delta \Sigma \approx (\Sigma_A - \Sigma_B)$. A ratio of $\alpha_h/\alpha_c$ of ten produces a duty cycle: $\tau_{high}/\tau_{FU}$ of approximately one tenth in agreement with observational constraints. For our standard case we take $\alpha_c = 10^{-4}$ and $\alpha_h = 10^{-3}$.

## 6. TIME DEPENDENT RESULTS

With the input mass flux in the range of $\dot{M}_{in} = (1 - 10) \times 10^{-6}$ M$_\odot$/ yr, we find for all $\alpha$'s a fully developed thermal outburst. The first annulus of the disk to cross the threshold of instability is near the inner edge of the disk. Its exact radius is determined by the details of the boundary condition (cf. §4.1.1.). With the no torque boundary condition, which causes a turn over in the first few values of $T_c$ and $\Sigma$, this annulus is generally at about 2 $R_*$. Once triggered, a weak heating front propagates inward from the ignition radius, heating the disk to some 10,000K and only partially ionizing hydrogen. When the front reaches the inner edge of the disk, radial effects no longer drain heat from both sides of the annulus and the thermal instability causes the local disk central temperature to rise sharply without a change in the surface density. The now fully developed ionization front propagates outward into the disk reaching a radius, $R_{limit}$, which depends upon $\dot{M}_{in}$ as discussed above in §5.2.. During the initial inward propagation of the front, mass flux through the innermost annulus ($\equiv \dot{M}_{acc}$) increases rapidly and then remains approximately constant ($\equiv \dot{M}_{FU}$) as the ionization front propagates outward. Once established, the front travels slowly out to its maximum excursion: $R_{limit}$, where radial effects: particularly the radial advection of heat, cause it to stabilize. The outburst lasts until the greatly enhanced mass flux: $\dot{M}_{FU} > \dot{M}_{in}$, drains the inner regions allowing the disk to cool below the recombination temperature of hydrogen and drop back onto the neutral lower branch.

In the following section (§6.1.) we trace the detailed evolution of our standard case model. We then vary our parameters (§ 6.2.) showing the sensitivity of our results to the magnitude of the input mass flux $\dot{M}_{in}$ (§ 6.2.1.), viscous efficiency $\alpha$ (§ 6.2.2.), inner boundary condition (§ 6.2.3.), and investigate the influence of our results to the details of our numerical implementation (§6.2.4.).

## 6.1. "Standard case" model

We discuss in detail the outburst of our standard model for which $\alpha_c = 10^{-4}$, $\alpha_h = 10^{-3}$, and $\dot{M}_{in} = 3 \times 10^{-6}$ $M_{\odot}$/ yr. This model fulfills our fitting criteria with peak surface temperatures in the disk remaining well over 6000K for most of the 140 years spent in outburst; time between successive outbursts is 900 years. The outburst mass flux: $\dot{M}_{FU} = 3 \times 10^{-5}$ $M_{\odot}$/ yr agrees well with observations of Fuor objects (§ 2.4.).

### 6.1.1. Onset of outburst

We begin the discussion immediately after outburst. Because the surface density at this time increases outwards nearly linearly ($\Sigma \approx \Sigma_B$) and the central temperature is essentially independent of radius, the mass flux in the inner regions has a strong radial dependence: $\dot{M} \sim \Sigma\nu \sim r^{5/2}$. As a consequence the surface density increases most strongly at the inner radii: $\partial\Sigma/\partial t \sim \Sigma\nu/r^2 \sim r^{1/2}$ (cf. Eq. 3), and the outburst is triggered first in the inner regions.

The radial distributions of $T_c$, $\Sigma$, $\mathcal{F}$, and $\dot{M}$ for the inner disk region ($r = 3$ to $10$ $R_{\odot}$) during the onset an outburst are shown in Figure 4 a, b, c, and d, respectively for the inner disk (out to 10 $R_{\odot}$). Fourteen curves are shown each separated by one year. Dashed lines in the surface density figure correspond to the critical surface densities: $\Sigma_A > \Sigma_B$. The dashed line in Figure 4c refers to the vertical thermal balance condition: $Q^- = Q^+$; positive $\mathcal{F}$ indicates local cooling, negative $\mathcal{F}$ indicates local heating. The dashed line in Figure 4d represents zero mass flux; above the line indicates mass moving radially outward, below indicates mass moving inward.

When the surface density at the ignition radius crosses the threshold of instability: $\Sigma > \Sigma_A$ (Fig. 4b), the ionization of hydrogen begins. The onset of hydrogen ionization increases the local viscosity and hence the local energy generation ($Q^+$) in two ways. Ionization increases the local sound speed and the local scale height leading to an increased viscosity; ionization also initiates the transition from $\alpha_c$ to $\alpha_h$ further increasing the local viscosity. At the same time that the magnitude of $Q^+$ is increasing, the disk enters a new opacity regime in which $Q^-$ becomes nearly independent of the central temperature. The disk's decreased ability to radiate is caused by the onset of convection and a rapidly increasing opacity which are both due to the ionization of hydrogen. The opacity effect is evident in the vertical structure results of $T_{eff}(T_c)$ in Figure 2b. For central temperatures between a few 1000 and 100,000 K, $T_{eff}$ (and thus $Q^-$) is nearly independent of $T_c$. The result of the combination of these two effects is a rapid local heating evidenced in Figure 4c where it can be seen that during the onset of ionization $Q^- << Q^+$. Once ionization begins, the triggered regions heat quickly at nearly constant effective temperature and surface density. Cooling due to the local surface density depletion (the PdV term in the energy equation) and the increased heat capacity during the ionization of hydrogen, however, prevent the unstable regions from heating to full ionization during the initial spread of the instability.

As pointed out in Lin et al. (1985), the zone which is first triggered into thermal instability behaves like an isolated high viscosity annulus. Mass spreads from this zone both inwards and outwards. The rapid diffusion of matter away from the ignition point causes the surface density to drop (Fig. 4b), but the annulus stays in the high state as long as $\Sigma > \Sigma_B$. A small amount of the mass from the partially ionized region in trying to spread outward slows the radially inward flowing matter and begins to pile up a modest surface density enhancement (the "snow plow" effect). There is not initially strong heating in the outward direction because radial advection is still bringing in cool matter from the outer regions and because of the effects of geometric dilution. The inward action of radial heat advection and the geometric effect, however, speed the inward spread of the instability and most of the displaced matter from the partially ionized region flows radially inward pushing successive annuli above their critical surface densities in an inward flowing "avalanche".

### 6.1.2. Propagation of ionization front

After the accretion of the inward flowing avalanche onto the central object which causes a modest mass flux spike in $\dot{M}_{acc}(t)$, the disk quickly attains a quasi–steady state structure in which the hot, ionized, inner disk region is separated from the cool, neutral, outer region by a slowly propagating ionization front. The evolution of $T_c$, $\Sigma$, $\mathcal{F}$, and $\dot{M}$ are shown for the disk out to 1/2 AU in Figure 5a-d during this period. Because the outward flowing "snow plow" is still quite close to the inner edge of the disk, the surface density in the fully ionized inner regions drops precipitously. Meanwhile, the outward mass flux from the high viscosity region has become sufficiently large such that $\dot{M}_{out} > \dot{M}_{in}$ and the sense of radial advection changes and now acts in the same sense as the thermal instability causing outward heat transport and propagation of the front. The surface density spike ($\Sigma > \Sigma_A$) which accompanies the ionization front now accelerates radially outward.

During the propagation phase of outburst, the disk becomes considerably thicker ($H/r \sim 0.3$) than it was in the quiescent phase ($H/r \sim 0.1$) such that radial transport of energy can comparable in magnitude to vertical transport. To examine the magnitude of these effects we examine in detail several of the terms of the energy equation (Eq. [5]). We define the radial dependence of $T_c$ such that $T_c \sim r^{-p}$, from which $\nu \sim T/\Omega \sim r^{\frac{3}{2}-p}$. Taking $\Gamma_3 = 5/3$ and using Eq. (4), the final terms of Eq. (5) can be written

$$Q_{rad} = \frac{\mu C_v}{R_g} \frac{\dot{M}}{2\pi} \left(\frac{c_s}{r}\right)^2 \left[\frac{\partial \ln T}{\partial \ln r} + \frac{2}{3}\left(\frac{\partial \ln \dot{M}}{\partial \ln r} - \frac{\partial \ln \Sigma}{\partial \ln r}\right)\right].$$

In regions where mass flux is approximately constant with radius (and it is evident from Fig. 5d that constant $\dot{M}(r)$ is set up quickly on the hot side of the front), $\Sigma \nu \sim r^0$ implies that $\Sigma \sim r^{p-\frac{3}{2}}$ and the bracketed part of the equation above can be expressed as $[1 - \frac{5}{3}p]$. If the temperature gradient decreases radially such that $p > \frac{3}{5}$, advection acts in the same sense as the mass flux. For this term to be comparable to the heating from local dissipation: $Q_{rad} \sim Q^+$, the temperature gradient must drop such that $p > \frac{3}{5}\left\{1 + \frac{3}{2}\frac{R_g}{\mu C_v}\left(\frac{\Omega r}{c_s}\right)^2\right\}$. In regions where either $C_v$ or $T_c$ is large, this condition can be met. Typical radial gradients during the propagation of the outburst are shown in Figure 5a.

At the leading edge of the ionization front, advection carries heat outwards and provides a strong local heating. In the region just interior to the ionization front, advection has the contrary effect which is to carry cool matter from partially ionized regions inward, providing a strong local cooling. One of the major impacts of the latter effect is to slow the local heating which allows the surface density to drop during the transition from low state to high state. The heat transfer in the radial direction is such that the final high state mass flux attained by a given annulus is considerably lower than it would have been had the annulus been isolated.

Radial radiative diffusion provides an effective tunneling which (in opposition to the advective term) flattens the radial temperature gradient in the inner low opacity region. We estimate this term using $T_c \sim r^{-p}$ as before and taking $F_s(rad)$ to be the surface flux in the radiative approximation. The radial radiative diffusion term can be simplified from Eq. (5):

$$Q_{rrd} = 8 \left(\frac{c_s}{\Omega r}\right)^2 p^2 (4-b) F_s (rad),$$

where $b$ is the temperature exponent of the opacity as in Appendix A.. Dividing by $Q^-$ = $2\mathrm{F}_s$ and taking $\mathrm{F}_s(rad) \approx \mathrm{F}_s$, we require $p^2 > \left(\frac{\Omega r}{c_s}\right)^2 \frac{1}{4(4-b)}$ for $Q_{rrd}$ to be significant. This term only has real solutions where $b < 4$. Because $b$ is negative or zero in the inner regions (Appendix A.), the term provides local heating behind the ionization front. When the opacity is high, however, as in the ionization front itself, the term is negligible.

The disk now consists of three distinct regions: The region radially outside of the ionization front is characterized by cool temperatures, neutral hydrogen, small radial gradients and local thermal balance. The transition region itself is characterized by high opacities, large radial gradients of $\Sigma$, $T_c$, and $\dot{M}$. Interior to this there is a slowly growing constant mass flux region (Fig. 5d) in which annuli are fully ionized. In this innermost region, steep temperature gradients ($p > \frac{3}{2}$ as in Fig. 5a) support a surface density which increases with radius (Fig. 5b). All constant mass flux regions are approaching vertical thermal balance: $Q^- = Q^+$ ($\mathcal{F} \to 0$; Fig. 5c), though not necessarily (as we discuss in the next section) on the upper branch of the equilibrium curve.

### 6.1.3. Decay from outburst

Unlike the standard CV case where the ionization front propagates all the way to the outer edge of the disk, the protostellar disk is sufficiently large that the instability is self–limiting. We now examine the situation the protostellar disk finds itself in as the ionization front propagates out into the disk; to aid the discussion, we refer to the equilibrium curves of $\dot{M}(\Sigma)$ for several radii shown in Figure 3b. If each annulus were completely isolated from its neighbors and could transport any amount of mass, at the onset of an outburst the mass flux through the initially triggered annulus would rise at constant surface density to the stable branch directly above $\Sigma_A$; for example, if the 5 $\mathrm{R}_\odot$ curve shown in Figure 3b were thermally isolated, when triggered it would rise on a thermal time scale directly to the upper branch at and begin transporting mass at $\approx 3 \times 10^{-4} \mathrm{M}_\odot / \mathrm{yr}$. Instead, the annulus drops significantly in surface density while increasing in mass flux due to the influence of neighboring low mass flux annuli. The result is that the outward mass flux spike, $\dot{M}_{out}$ ($\approx 10^{-4} \mathrm{M}_\odot / \mathrm{yr}$), and inward high state mass flux, $\dot{M}_{FU}$ ($\approx 3 \times 10^{-5} \mathrm{M}_\odot / \mathrm{yr}$) are considerably less than the expected isolated high state value of $3 \times 10^{-4} \mathrm{M}_\odot / \mathrm{yr}$ (for the purposes of vertical thermal balance, it does not matter whether a given mass flux is inward or outward). Once these upper branch mass fluxes are established they provide an inner boundary condition which influences the propagation of the ionization front and so persist throughout the outburst.

In Figure 6, the details of the path of $T_{eff}(\Sigma)$ (solid line) taken by the 20 $R_\odot$ annulus during a complete outburst cycle is traced (evolution proceeds in a counter–clockwise direction) along with its appropriate equilibrium curve (dashed line). The annulus is first triggered by the strong outward mass flux pulse $\dot{M}_{out}$ which pushes $\Sigma$ above the local $\Sigma_A$ indicated in the figure as a progression up and to the right of the equilibrium curve. The annulus then continues to heat while losing surface density to adjacent annuli as indicated by a progression up and to the left. For the innermost annuli, as for the uppermost arrow in Fig. 3b for 5 and 10 $R_\odot$, the path the annulus takes leading to $\dot{M}_{FU}$ intercepts the equilibrium curve ($\mathcal{F}=0$) on the stable upper branch. Beyond some 10 $R_\odot$, however, $\dot{M}_{FU}$ (in this example $3 \times 10^{-5}$ $M_\odot/$ yr) has no stable upper branch. The annulus is therefore pushed above $\Sigma_A$, but in evolving to $\dot{M}_{FU}$ intercepts the equilibrium curve in an unstable region. This is shown explicitly in Fig. 6 for the annulus at 20 $R_\odot$. Passing through the $\mathcal{F}=0$ equilibrium curve, thermal evolution proceeds only slowly as radial effects easily balance the $Q^- - Q^+$ term of the energy equation and $dT/dt\approx0$. This effect is enhanced by the details of our vertical structure model which result in a narrow intermediate stable branch at about $T_c = 10^4$ K. As a result, the disk interior to the ionization front stabilizes on the thermal equilibrium curve for $\dot{M}_{FU}$ even out to radii where such an equilibrium is unstable.

The front can only be self–sustaining out to the limit radius $R_{limit}$ where $\Sigma_A$ equals $\Sigma(\dot{M}_{in})$ (as discussed in §5.2.); beyond this radius $\dot{M}_{in}$ provides a surface density which is on the stable low branch at a value below $\Sigma_A$. The front can continue to propagate to some degree, but at the radius where the front can no longer increase $\Sigma$ above the critical $\Sigma_A$, propagation ceases and the ionization front stalls. The radial dependence of $T_c$, $\Sigma$, $\mathcal{F}$, and $\dot{M}$ are shown during decay in Figure 7a-d. For the standard case model under discussion, $R_{limit} = 20$ $R_\odot$ and the ionization front propagates just slightly beyond this radius. At the outer radius where the front stalls, the large outward mass flux, $\dot{M}_{out}$, and large inward mass flux, $\dot{M}_{FU}$, combine to cause the surface density at this annulus to drop. Because the annuli immediately interior to $R_{limit}$ have not attained the upper stable branch, the surface density does not need to drop below $\Sigma_B$ for the cooling wave to begin its inward transition. Once triggered, the ionization wave retreats rapidly dropping an a local thermal time scale (the vertical portion of Fig. 6 for $r = 20$ $R_\odot$) directly down to the lower stable branch. A pause in the temperature decrease is evident in Figure 7a due to the increase of specific heat during the recombination of hydrogen. This disk is left strongly out of viscous equilibrium (note the strong radial dependence of $\dot{M}(r)$ immediately after outburst in Fig. 7d) and on the low state viscous time scale slowly begins accumulating matter for another outburst.

## 6.2. Variation of parameters

We now investigate the dependence of outburst properties on a variety of parameters. Light curves: $\dot{M}_{acc}(t)$ and $L_{bol}(t)$, are all shown in Figure 8. The standard case model for which $\dot{M}_{in} = 3 \times 10^{-6}$ $M_\odot$/yr, $\alpha_c = 10^{-4}$, and $\alpha_h = 10^{-3}$ is displayed in the right hand column of Figure 8a. $\dot{M}_{acc}$ corresponds to the mass flowing through the innermost annulus. $L_{bol}$ is given by the sum of $\sigma T_{eff}^4$ times the area of each annulus from the inner edge of the disk to the outer edge (1/2 AU) well beyond the outbursting region of the disk. This luminosity is strictly from *one side* of the disk[4] and does not include any boundary layer or central object radiation.

### 6.2.1. Dependence upon $\dot{M}_{in}$

We first investigate the dependence of our models on input mass flux, $\dot{M}_{in}$ for our standard viscosity: $\alpha_c = 10^{-4}$ and $\alpha_h = 10^{-3}$. Time scales, outburst mass fluxes and luminosities, and peak temperatures are summarized in Table 2 for four values of $\dot{M}_{in}$. We define $\tau_{rise}$ to be the time it takes for the system to rise to peak light after passing through $\dot{M}_{crit}$, $\tau_{high}$ to be the time between the rising and falling of $\dot{M}_{acc}$ past $\dot{M}_{crit}$, and $\tau_{FU}$ to be the time for the cycle to repeat. The columns $\dot{M}_{FU}$, $L_{bol}$, and $T_{eff}$ are peak values during outburst neglecting any initial spike. Light curves are illustrated for two values of $\dot{M}_{in}$: $1 \times 10^{-6}$ and $3 \times 10^{-6}$ $M_\odot$/yr in the right (R) and left (L) panels of Figure 8a respectively.

For our standard $\alpha$: $\alpha_c = 10^{-4}$ and $\alpha_h = 10^{-3}$, we are able to fit within observational constraints for the range of input mass fluxes $\dot{M}_{in} = (1 - 10) \times 10^{-6}$. Outburst surface temperatures are well matched in this mass flux range. Time spent in the high state, $\tau_{high}$, ranges from 85 to 250 years; time between successive outbursts, $\tau_{FU}$, ranges from 800 to 1200 years and thus fit our time scale constraints (§ 2.4.). As $\dot{M}_{in}$ increases, all time scales and temperatures increase. Maximum central temperatures in the high state are $(1 - 2) \times 10^5$ K with this particular viscosity. High state mass fluxes through the inner zone: $\dot{M}_{FU} = (1 - 5) \times 10^{-5}$ $M_\odot$/yr, are comparable to modeled FU Orionis values (Table 1). Luminosities in the high state also agree favorably with observations of FU Orionis objects. For modeled outbursts the duty cycle ($\tau_{high}/\tau_{FU}$) increases with $\dot{M}_{in}$; for $10^{-6}$ $M_\odot$/yr, $\tau_{high}/\tau_{FU} \approx 0.1$ while for $10^{-5}$ $M_\odot$/yr, $\tau_{high}/\tau_{FU} \approx 0.2$. The lower mass flux rise times produced by these "inside–out" self–regulated outbursts fit the observations of outbursting Fuor V1515 Cyg; all are too long to fit either FU Ori or V1057 Cyg. It has been suggested that these more rapid rise times might result from an "outside–in" outburst which begins near the outer edge of the unstable region and propagates inward (Clarke et al. 1990; Hartmann et al. 1993). We shall present in a subsequent paper detailed spectral fits where such outbursts may be triggered by modest perturbations in the outer regions of unstable disks (Bell et al. 1994).

---

[4]Luminosity must therefore be multiplied by two to compare to the derived isotropic values in Table 1.

### 6.2.2. Dependence upon $\alpha$

Sensitivity of derived time scales on the magnitude of $\alpha$ are tested with computation of time dependent radial evolution models for values of $\alpha$ from $10^{-1}$ and $10^{-4}$ and with $\alpha_{\rm h}/\alpha_{\rm c}$ ratios ranging from one to ten. This extensive numerical exploration of parameter space shows that many different values of $\alpha$ result in disks subject to thermal outbursts. In the case where the ratio $\alpha_{\rm h}/\alpha_{\rm c}$ is *less* than one, the unstable portion of the equilibrium curve is reduced in extent and outbursts are not seen. Increasing $\alpha_{\rm h}/\alpha_{\rm c}$ above one has the effect of increasing the length of the unstable portion of the cuvre and thus enhancing the disk outburst mechanism. From the time scale discussions above (§5.3.) it is argued that the ratio of time spent in the low state to time spent in the high state depends upon the ratio of viscous time scales. We observe that models with $\alpha_{\rm h}/\alpha_{\rm c} \approx 10$ result in duty cycles in good agreement with observational statistics. As expected from the equilibrium curve analysis, the value of $\dot{\rm M}_{\rm crit}$ is found to be *independent of the value of* $\alpha$ and in all cases is equal to $5 \times 10^{-7}$ M$_\odot$/ yr.

With $\alpha_{\rm h}/\alpha_{\rm c}$ set to ten, outbursts are found to occur for all magnitudes of $\alpha_{\rm c}$ tested. Larger values of $\alpha$ result in shorter outburst time scales because of the increased diffusivity of matter in the disk. We therefore use the magnitude of the viscosity as a free parameter in our time scale fitting. To illustrate the dependence of these models on the magnitude of $\alpha$, we present in Figure 8b as well as in Table 2 results from two test cases. In both test cases $\dot{\rm M}_{\rm in} = 3 \times 10^{-6}$ M$_\odot$/ yr. In Figure 8bL we display the result of dropping the magnitude of $\alpha_{\rm h}$ by a third to $3 \times 10^{-4}$ while keeping $\alpha_{\rm c}$ equal to $10^{-3}$. The resulting system spends spends a third of its time in outburst: duty cycle $\approx 0.3$ where the increased $\tau_{\rm high}$ is a direct result of the lower $\alpha_{\rm h}$. The high state effective temperature and luminosity (Table 2) are significantly lower than in the standard case as the result of the combination of lower $\dot{\rm M}_{\rm FU}$ and lower $\alpha$. In Figure 8bR, we display results of a model for which both $\alpha$'s are ten times the standard such that $\alpha_{\rm c} = 10^{-3}$ and $\alpha_{\rm h} = 10^{-2}$. The larger $\alpha$ results in time scales which are too short to reproduce observed FU Orionis time scales and surface temperatures which are too high. Despite the considerably shorter time scale in this high $\alpha$ model, the $\tau_{\rm rise}$ of 6 yrs (Table 2) is still too long to reproduce either the FU Ori or V1057 Cyg outburst.

### 6.2.3. Variation of inner boundary condition

In Figure 8c, we produce models calculated using (L) a different inner boundary condition and (R) a larger inner disk radius. These models make use of the standard viscosity: $\alpha_c = 10^{-4}$, $\alpha_h = 10^{-3}$, and standard input mass flux: $\dot{M}_{in} = 3 \times 10^{-6}$ $M_\odot$/yr. The details of the inner boundary condition are expected to have a significant impact on conditions in the inner regions of the disk (e.g., Pringle 1981). For example, the inclusion of a more realistic photosphere may significantly modify the accretion process during the initial phase of the outburst when the inward propagating avalanche is accreted onto the central object. Although it is our intention here to focus on the propagation of the ionization front through the protostellar disk, we explore the robustness of our results by altering the inner boundary condition in several test cases.

The standard case makes use of the boundary condition for which $\Sigma_0 = 0$; this condition is the "fall into a hole" condition commonly employed in accretion disk research. The condition can be justified by the observation that low mass YSOs are generally rotating well below break–up speed: $\Omega_* << \Omega_K(R_*)$, and assumes that the central object must be decoupled in some way from its inner disk. Figure 8cL shows the effects of using a radially constant mass flux inner boundary condition: $(\Sigma\nu)_0 = \Sigma_1\nu_1$, which is equivalent to the opposite extreme in which $\Omega_* = \Omega_K(R_*)$. The time scales derived with this boundary condition are essentially identical to the standard case. The primary differences of this model from our standard case model (Fig. 8aR) occur because $\Sigma$ and $T_c$ do not curve over in the inner few zones and result in a higher $T_{eff}$ at the inner edge and hence higher $L_{bol}$. These values are so high as to cause the model to fall outside of our constraints; lowering $T_{eff}$ would require a lower value of $\alpha_h$ which would then lengthen the time scales so as to again put the model outside of our constraints. We also note the unacceptably low $\dot{M}_{crit}$ of $2 \times 10^{-8}$ $M_\odot$/yr.

Another assumption we have made about the inner boundary condition is that the inner disk extends all the way down to the stellar surface. There is no *a priori* reason for expecting this to be the case; indeed some calculations have suggested that the inner disks of T Tauri objects may be truncated in the presence of 1 kG stellar magnetic field (Königl 1991). We therefore calculate a test case in which we use an inner radius of 6 $R_\odot$: twice our standard value. Figure 8cR shows the results of this model. Although the essential time scales are not dramatically affected, the triggering mass flux: $\dot{M}_{crit}$, is considerably higher in this case than for the standard boundary condition. This can be understood in terms of the strong radial dependence of the equilibrium curves discussed above (§5.2. and Fig. 3). That boundary layer mass fluxes among T Tauri objects are generally not seen to be this high (there are exceptions, for example DR Tau seems to show not only an unusually large mass flux but also a significant inner hole in its disk [Kenyon et al. 1994]) suggests that few *actively accreting* low mass YSOs have significant inner disk holes. We do not expect thermal disk outbursts of the kind under consideration here in systems for which the protostellar accretion disk has a hole as large as or larger than $R_{limit}$.

We therefore summarize that altering our inner boundary condition from the standard case has no significant impact on derived time scales but may have detrimental impact on goodness of fit in $T_{eff}$, $L_{bol}$, and $\dot{M}_{crit}$.

### 6.2.4. Variation of numerical resolution

Finally, in Figure 8d, we produce results from parameter variations involving the size of our time step, and the number of radial zones. These models again make use of the standard viscosity: $\alpha_c = 10^{-4}$, $\alpha_h = 10^{-3}$, and standard input mass flux: $\dot{M}_{in} = 3 \times 10^{-6}$ $M_\odot$/yr. The variations do not result in significant changes to time scales and confirm that our results are not dependent upon the mechanical details of the numerical model.

Light curves are shown in Figure 8dR resulting from a reduction of the time step to the shortest photon zone–crossing time and in Figure 8dL resulting from an increase in the number of radial zones is from 72 to 100. The models took between three and ten times longer to compute than, and resulted in results essentially identical to, the standard case.

## 7. SUMMARY

Our primary intention in this contribution has been to put constraints on the magnitude of protostellar disk viscosity through the modeling of FU Orionis outbursts as self–regulated accretion events in protostellar accretion disks. This outburst scenario was initially proposed to explain qualitatively similar outbursts seen in cataclysmic variables. Adaptations in the current work include the effects of (1) convection, (2) radial transport of heat, (3) updated opacities, and (4) the effects of a local departure from vertical thermal balance. We find that with a 1 $M_\odot$, 3 $R_\odot$ central object, time scales and surface temperatures of FU Orionis outbursts can be matched using the *ad hoc* $\alpha$ viscosity law with values of $\alpha = 10^{-4}$ in the cool, neutral hydrogen regions of the nebula and $10^{-3}$ in the hot, ionized regions and input mass fluxes in the range of $(1 - 10) \times 10^{-6}$ $M_\odot$/yr. With these parameters, the protostellar disk is regulated such that brief periods ($\approx 100$ yrs) of high central object mass flux ($10^{-5}$ $M_\odot$/yr) alternate with long periods ($\approx 1000$ yrs) of low mass flux ($10^{-7}$ $M_\odot$/yr). Although the exact magnitude of $\alpha$ can not be taken any more literally than its *ad hoc* functional form warrants, the low magnitude of the viscosity it implies combined with the large predicted mass infall rates suggest surface densities and midplane temperatures considerably larger than generally thought to occur in protostellar disks. Masses of these modeled disks are such that the effects of self–gravity may become significant in the disk as close to the central object as one AU.

General features of modeled outbursts include:

1. Disks transporting mass at a rate greater than the critical value $\dot{M}_{crit} = 5 \times 10^{-7}$ M$_\odot$/yr are subject to self–regulated, FU Orionis type, repetitive accretion events. The value of $\dot{M}_{crit}$ is independent of $\alpha$.

2. Triggering of the outburst occurs due to the sudden changes in opacity and to the onset of convection which occur as a result of the ionization of hydrogen. The initiation of an outburst occurs at that radius for which the surface density increases above the local critical surface density $\Sigma_A$.

3. Self–regulated outbursts are generally initiated near the inner edge of the disk region resulting in an "inside–out" progression of the ionization front.

4. The ionization front propagates to a radius which is determined primarily by the magnitudes of $\dot{M}_{in}$ and $M_*$. All outbursts investigated here are limited to within 1/4 AU; beyond this region, during both FU Orionis and T Tauri phases, mass is stably transported at a rate equal to $\dot{M}_{in}$.

5. Between outbursts, the mass flux through the inner annuli of modeled disks can be as little as 1% of the mass flux at 1/2 AU; during outburst, the central mass flux can be as much as ten times the outer disk mass flux.

6. Time scales associated with the outburst are highly dependent on the magnitude of the viscosity and are dependent to a lesser degree on the value of input mass flux, $\dot{M}_{in}$.

A prediction of this model is that outbursting FU Ori objects should have a transition radius at about 1/4 AU outside of which mass flux is considerably smaller than what is estimated to be passing through the innermost annuli. This effect should be detectable spectroscopically as a lower than expected flux at wavelengths of a few microns. Current estimates suggest that this departure from the steady state models of Hartman and Kenyon may be detectable but is not a dominant effect (Bell et al. 1994).

## 8. DISCUSSION

Above a certain infall mass flux, the disk is expected to attain a state where the ionization front is stalled at some radius resulting in a disk in constant outburst. Clarke et al. (1899) calculate a case in which the front marking the transition from neutral to ionized hydrogen is stalled at $30R_\odot$ with $\dot{M}_{in} = 10^{-4}$ $M_\odot/$ yr and $\alpha = 10^{-3}$. We observe this phenomenon in our models in the trend that the duty cycle lengthens with increasing $\dot{M}_{in}$ for a given $\alpha$. In the disks of cataclysmic variables there is a mass flux above which the entire disk is sufficiently hot to remain fully ionized. For the much larger protostellar disks, the mass flux of stabilization should occur well below the mass flux which would put the entire disk in outburst. The mass flux to produce a stable ionization front is likely to be well above typical molecular cloud core infall rates for low mass systems but may be important for the higher mass Herbig Ae/Be systems.

The outbursting mechanism herein examined requires a steady flux of mass onto the outer accretion disk of at least some fraction of YSOs. We hypothesize that such a steady flow is produced in most Fuor systems by remnant infall from a spherical cloud core. Shu, Adams, & Lizano (1987) estimate accretion from a spherical core independent of central condensation mass or initial density of $\dot{M} \approx a^3/G$ where $a$ is the sound speed present in the cloud before collapse. Adams, Lada, & Shu (1987) derive $a$ from spectral fits of embedded protostars in the range of 0.20 to 0.35 km/s. These values lead to mass fluxes of $2 \times 10^{-6}$ to $1 \times 10^{-5}$ $M_\odot/$ yr or time scales of 1 $M_\odot$ objects in the embedded phase $M_\odot/\dot{M}_{in}$ of a few hundred thousand years. Strom et al. (1993) estimate survival times of disks optically thick out to 1 AU to be $(1-10) \times 10^6$ yrs. If a system is subject to FU Orionis outbursts primarily during its embedded phase, we thus expect some 10% of all observed low mass YSOs to be sufficiently young to be subject to Fuor outbursts; the value of 10% was used in §2.4. in our statistical derivation of the constraint $\tau_{FU}$. That the large low state disk luminosities we derive (1–6 $L_\odot$) are generally not seen in T Tauri objects ($L_{IRE} \approx 1$ $L_\odot$) also supports the idea that objects subject to FU Orionis outburst are likely to be heavily embedded and therefore young protostars.

The inference that FU Orionis outbursts occur early in the accretion disk stage is given weight by the observation that FU Orionis objects are observed to be heavily embedded. Fuors are so commonly associated with arc shaped reflection nebulosity (see Table 1) that the existence of such nebulosity was originally a defining characteristic (Herbig 1977). Further, Kenyon & Hartmann (1991) show that although optical spectra and near IR excesses are well matched by accretion disk models, longer wavelength observations of Fuors ($\lambda > 10\mu m$) show significant excess above what is predicted by the simple accretion disk scenario. They model observations of the post-outburst fading of this long wavelength

"excess" excess in Fuor V1057 Cyg as arising from light reprocessed by a spherically infalling cloud with a wind driven polar hole which contacts the disk at a distance of 6 AU from the central object. Density considerations lead them to estimate a spherical infall rate for V1057 Cyg of $4 \times 10^{-6}$ M$_\odot$/yr consistent with mass fluxes derived in this work.

We wish to acknowledge helpful conversations with P. H. Bodenheimer and L. Hartmann. K. R. B. would like to acknowledge support from a Sigma Xi award and from NASA training grant NGT–50665. Part of this work was conducted under the auspices of a special NASA astrophysics theory program that supports a Joint Center for Star Formation Studies at NASA Ames Research Center, the University of California at Berkeley, and the University of California at Santa Cruz and was also supported in part by NASA Origins grant NAGW–2306.

## A.  Opacity

Because the details of the opacity law have a profound effect on the propagation of the thermal instability, we have modified the Lin & Papaloizou (1985) analytic Alexander / Cox / Stewart opacities taking into account the recent 1 – 3000K revisions suggested by Alexander et al. (1989). In particular we define a frequency averaged opacity law in units of $cm^2/gm$:

$$\kappa = \kappa_i \rho^a T^b,$$

over eight regions in order of ascending temperature:

| | | $\kappa_i=$ | $a=$ | $b=$ |
|---|---|---|---|---|
| 1. | Ice grains: | $2 \times 10^{-4}$ | 0 | 2, |
| 2. | Evaporation of ice grains: | $2 \times 10^{16}$ | 0 | -7, |
| 3. | Metal grains: | 0.1 | 0 | 1/2, |
| 4. | Evaporation of metal grains: | $2 \times 10^{81}$ | 1 | -24, |
| 5. | Molecules: | $10^{-8}$ | 2/3 | 3, |
| 6. | H- scattering: | $10^{-36}$ | 1/3 | 10, |
| 7. | Bound–free and free–free: | $1.5 \times 10^{20}$ | 1 | -5/2, |
| 8. | Electron scattering: | 0.348 | 0 | 0. |

Although the listed processes are the dominant effects in each temperature regime, a priority is given to a good fit to tabulated values rather than to explicit temperature dependence as defined from atomic principles. Transitions occur where $\kappa_i = \kappa_{i+1}$ and are smoothed following the method of Lin & Papaloizou (1985). Opacities using these analytic laws are compared to tabulated Alexander et al. values for densities typical of the inner protosolar nebula in Figure 9a. Equilibrium curves calculated with the old opacities are compared to results using the new opacities in Figure 9b. With the older opacities, a protostellar disk would be expected to be stable up to mass fluxes well over $10^{-6}$ $M_\odot$/ yr.

Table 1: Timescales and Modeled Properties of Known and Suspected Fuors

| Object | Beginning of outburst | $\tau_{rise}$ yrs | $\Delta M$ | $L_{bol}$ $L_\odot$ | $\dot{M}_*M$ $M_\odot^2/yr$ | Arc? | HH? |
|---|---|---|---|---|---|---|---|
| FU Ori | 1936 | 0.5–1 | 6 pg | 490 | $10^{-4}$ | yes | |
| V1057 Cyg | 1969 | 1 | 5.5 pg | 350 | $10^{-4}$ | yes | yes |
| V1515 Cyg | 1920–1945 | $\approx 20$ | 4 pg | 200 | $10^{-4}$ | yes | |
| V1735 Cyg[a] | 1957–1965 | <8? | >5 R | >75 | | yes | |
| V346 Nor[b] | 1978–1980 | 3–5 | >2.5 J | 288 | | yes | yes |
| RNO 1B/1C | 1978–1990 | < 12? | > 3 R | 1000(B+C) | | yes | |
| BBW 76 | <1956 | | | 550 | | yes | |
| Z CMa | <1860? | | | 1000–3000 | $10^{-3}$ | | yes |
| L1551 | | | | >40 | $10^{-5}$ | | yes |
| Parsamyan 21 | | | | | | yes | yes |

References: Cohen & Schwartz 1987; Eislöffel, Hessman, & Mundt 1992; Elias 1978; Graham & Frögel 1985; Hartmann 1991; Hartmann, Kenyon, & Hartigan 1993; Herbig 1977; Hessman et al. 1991; Kenyon, Hartmann, & Hewett 1988; Kenyon et al. 1993; Koresko et al. 1991; Poetzel, Mundt, & Ray 1990; Reipurth 1989; Rodriguez, Hartmann, & Chavira 1990; Staude & Neckel 1991, 1992.

[a]V1735 Cyg = Elias 12
[b]V346 Nor = object in HH57

Table 2: Results of Outburst Models

| $\alpha_c$ | $\alpha_h$ | $\dot{M}_{in}$ $10^{-6}\frac{M_\odot}{yr}$ | $\tau_{rise}$ yrs | $\tau_{high}$ yrs | $\tau_{FU}$ yrs | $\dot{M}_{FU}$ $10^{-6}\frac{M_\odot}{yr}$ | $L_{bol}$[a] $L_\odot$ | $T_{eff}$[b] K |
|---|---|---|---|---|---|---|---|---|
| $10^{-4}$ | $10^{-3}$ | 1 | 25 | 85 | 780 | 10 | 14 | 5600 |
| | | 3 | 50 | 140 | 900 | 30 | 35 | 6800 |
| | | 5 | 60 | 170 | 1050 | 40 | 60 | 7300 |
| | | 10 | 80 | 250 | 1150 | 50 | 85 | 8000 |
| $10^{-4}$ | $3 \times 10^{-4}$ | 3 | 90 | 270 | 700 | 7 | 11 | 5000 |
| $10^{-3}$ | $10^{-2}$ | 3 | 6 | 12 | 160 | 40 | 65 | 8500 |

[a]The bolometric luminosity given is the peak during outburst and includes radiation from only one surface of the disk.
[b]Temperature is the maximum value during outburst.

### Figure captions

Figure 1. Full non–VTB vertical structure grid $\Sigma(T_c, \mathcal{F})$ for use in time dependent diffusion models. Surface plot above shows dependence of $\Sigma$ on $T_c$ and $\mathcal{F}$ at 0.1 AU with $\alpha = 10^{-4}$. The disk is thermally unstable in regions where $\partial\Sigma/\partial T_c < 0$. A cross section of this plot taken at the VTB condition $Q^- = Q^+$ shown below indicates that the thermal instability occurs in regions where the central temperature falls between $10^4$ and $10^5$ K.

Figure 2. Equilibrium curves for four values of $\alpha$: $10^{-1}$, $10^{-2}$, $10^{-3}$, and $10^{-4}$ calculated at 4.1 R$_\odot$. Solid curves are results of convective VS models for all four $\alpha$'s and dashed curves are results of radiative VS models for highest and lowest $\alpha$'s. (a) $T_{eff}(\Sigma)$ and $\dot{M}(\Sigma)$ ($T_{eff}$ and $\dot{M}$ are uniquely related for a single radius through Eq. [2]). The critical surface density, $\Sigma_A$, for $\alpha = 10^{-4}$ and the critical mass flux, $\dot{M}_{crit}$, valid for all $\alpha$'s are indicated. Note the strong dependence of $\dot{M}_{crit}$ on $\alpha$ in the radiative models. (b) $T_{eff}(T_c)$ and $\dot{M}(T_c)$. Surface temperature is relatively independent of central temperature between $10^4$ and $10^5$ K. (c) $T_c(\Sigma)$. For convective models, central temperature depends upon the magnitude of the viscosity. (d) $(H_d/r)(\dot{M})$. The disk thickens considerably during the ionization of hydrogen.

Figure 3. Equilibrium curves for standard viscosity: $\alpha_c = 10^{-4}$, $\alpha_h = 10^{-3}$, at five radial points: $r = 5$, 10, 20, 40, and 80 R$_\odot$. (a) $T_{eff}(\Sigma)$. The critical effective temperature is $\approx 2000$K relatively independent of $r$. (b) $\dot{M}(\Sigma)$. Radial propagation of front can be estimated from input mass flux. Representative mass fluxes shown by arrows: stable low state mass flux: $10^{-7}$ M$_\odot$/ yr, input mass flux which results in outbursting model: $3 \times 10^{-6}$ M$_\odot$/ yr, and mass flux in inner disk during outburst: $3 \times 10^{-5}$ M$_\odot$/ yr.

Figure 4. Onset of outburst: time evolution of radial distributions during time dependent diffusion calculation of the "standard model" for which $\dot{M}_{in} = 3 \times 10^{-6}$ M$_\odot$/ yr,

$\alpha_c = 10^{-4}$, and $\alpha_h = 10^{-3}$. Successive snapshots (14) are each separated by one year for the inner disk region ($r = 3$–$10 R_\odot$). (a) $T_c(r)$. Inner regions heat during outburst. (b) $\Sigma(r)$. Upper and lower dashed lines indicate $\Sigma_A(r)$ and $\Sigma_B(r)$ respectively. Outburst initiated at radius where $\Sigma > \Sigma_A$. (c) $\mathcal{F}(r)$. Fractionalized heat imbalance: $\mathcal{F} = (Q^- - Q^+)/(Q^- + Q^+)$. Dashed line indicates condition for vertical thermal balance: $Q^- = Q^+$. Negative values indicate local thermal heating, positive values indicate cooling. (d) $\dot{M}(r)$. Dashed line indicates zero mass flux; below dashed line indicates matter flowing inward, above indicates matter flowing outward.

Figure 5. Propagation of outburst (as in Fig. 4): time evolution of (a) $T_c(r)$, (b) $\Sigma(r)$, (c) $\mathcal{F}(r)$, and (d) $\dot{M}(r)$. Successive snapshots (10) are each separated by intervals of three years. The ionization front is accompanied outward by a spike in the surface density (b) and by a strong outward mass flux pulse (d). Negative values of $\mathcal{F}$ (c) indicate strong local heating.

Figure 6. $T_{eff}(\Sigma)$. Path taken in parameter space during one full cycle of the standard model by annulus at 20 $R_\odot$; evolution proceeds counter–clockwise. Dashed line indicates equilibrium curve. Local heating occurs to the right of the line ($\mathcal{F} < 0$) and cooling to the left ($\mathcal{F} > 0$). In an isolated system the temperature is expected to rise at constant surface density to the upper branch; in this system, radial effects lead to more complicated behavior. System leaves equilibrium curve and moves to the right and up during passage of surface density spike (Fig. 5b). Depletion of the surface density during heating then results in movement up and to left preventing direct rise to upper stable branch. The system evolves slowly (on time scales of decades) through the equilibrium curve then quickly (years) drops back to stable lower branch during retreat of ionization front. During quiescence matter accumulates on a viscous time scale (centuries) and the system moves back up lower branch of equilibrium curve as conditions evolve toward the onset of the next outburst.

Figure 7. Decay from outburst (as in Fig. 4): time evolution of (a) $T_c(r)$, (b) $\Sigma(r)$, (c) $\mathcal{F}(r)$, and (d) $\dot{M}(r)$. Successive snapshots (13) are separated by intervals of three years. Decay begins at outer edge of the unstable region and propagates inward. Outer disk: $r > 30$ $R_\odot$, unaffected by outburst in inner disk. Note strong radial dependence of $\dot{M}$ (d) in inner disk immediately after outburst.

Figure 8. $\dot{M}_{acc}(t)$ and $L_{bol}(t)$: Parameter variations. Note that $L_{bol}$ is the luminosity from *one surface* of the disk. (a) Light curves for two different input mass fluxes using the standard $\alpha$: $\alpha_c = 10^{-4}$, $\alpha_h = 10^{-3}$. aL & aR: $\dot{M}_{in} = 1$ and $3 \times 10^{-6}$ $M_\odot$/ yr respectively; **aR is the "standard case"**. $\tau_{high}$ and duty cycle lengthen with $\dot{M}_{in}$; $\tau_{FU}$ relatively unaffected. (b) Light curves for constant $\dot{M}_{in} = 3 \times 10^{-6}$ $M_\odot$/ yr and variable $\alpha$. bL Hot $\alpha$ smaller than standard case: $\alpha_h = 3 \times 10^{-4}$. bR Both $\alpha$'s larger than standard case: $\alpha_c = 10^{-3}$ and $\alpha_h$

$= 10^{-2}$. Both $\alpha$ variations result in unacceptable time scales. All four remaining models use standard parameters: $\alpha_c = 10^{-4}$, $\alpha_h = 10^{-3}$, and $\dot{M}_{in} = 3 \times 10^{-6}$ $M_\odot/$ yr. (c) Light curves for inner boundary condition variations. cL Zero radial gradient in $\dot{M}$ inner boundary condition. Small $\dot{M}_{acc}$ before and large $L_{bol}$ during outburst reflect larger temperatures at inner edge of disk. cR Large inner disk radius of 6 $R_\odot$. Large $\dot{M}_{acc}$ before and small $L_{bol}$ during outburst reflect lower temperatures near inner edge of disk. (d) Variations in grid. dL Time step limited by photon diffusion zone crossing time. dR Larger number of radial grid points. Time scales comparable to standard case.

Figure 9. (a) $\kappa(T)$ for $\rho = 10^{-5,-6,-7,-8,-9}$ gm/cm$^3$. Solid lines give Alexander et al. (1989) tables; dashed lines give analytic approximation used in all calculations. Steeply rising region between 2000 and 10,000K gives rise to thermal instability. (b) $T_{eff}(\Sigma)$. Effects of opacity on thermal equilibrium curves. Solid lines give VS results for four $\alpha$'s ($10^{-1}$, $10^{-2}$, $10^{-3}$, and $10^{-4}$) using new analytic opacities; dashed lines make use of previous analytic opacities (Lin & Papaloizou 1985).